\documentclass[aps,pra, twocolumn,groupedaddress]{revtex4-1}

\usepackage{graphicx}
\usepackage{amsmath}
\usepackage{epsfig}
\usepackage{dcolumn}
\usepackage{bm}
\usepackage{color}
\usepackage{epstopdf}

\newcommand{\nc}{\newcommand}
\nc{\eq}{\begin{equation}}
\nc{\eeq}{\end{equation}}
\nc{\eqa}{\begin{eqnarray}}
\nc{\eeqa}{\end{eqnarray}}
\nc{\ar}{\begin{array}}
\nc{\ear}{\end{array}}
\nc{\bfig}{\begin{figure}}
\nc{\efig}{\end{figure}}
\nc{\dg}{\dagger}
\nc{\sx}{\sigma_x}
\nc{\sy}{\sigma_y}
\nc{\sz}{\sigma_z}
\nc{\spl}{\sigma_+}
\nc{\sm}{\sigma_-}
\nc{\nn}{\nonumber}
\nc{\noi}{\noindent}
\nc{\adg}{a^{\dg}}
\nc{\kvec}{\mathbf{k}}
\nc{\xvec}{\mathbf{x}}

\def\bra#1{\mathinner{\langle{#1}|}}
\def\ket#1{\mathinner{|{#1}\rangle}}

\begin{document}


\title{Non-Markovian probes in ultracold gases}

\author{P.~Haikka$^{1}$, S.~McEndoo$^{1,2}$, S.~Maniscalco$^{1,2}$}

\affiliation{$^1$ Turku Center for Quantum Physics, Department of
Physics and Astronomy, University of Turku, FIN20014, Turku,
Finland} 
 \affiliation{$^2$ SUPA,
EPS/Physics, Heriot-Watt University, Edinburgh, EH14 4AS, United Kingdom,}


\date{\today}

\begin{abstract}
We present a detailed investigation of the dynamics of two physically different qubit models, dephasing under the effect of an ultracold atomic gas in a Bose-Einstein condensed (BEC) state. We study the robustness of each qubit probe against environmental noise; even though the two models appear very similar at a first glance, we demonstrate that they decohere in a strikingly different way. This result holds significance for studies of reservoir engineering as well as for using the qubits as quantum probes of the ultracold gas. For each model we study whether and when, upon suitable manipulation of the BEC, the dynamics of the qubit can be described by a (non-)Markovian process and consider the the effect of thermal fluctuations on the qubit dynamics. Finally, we provide an intuitive explanation for the phenomena we observe in terms of the spectral density function of the environment. \end{abstract}


\maketitle

\section{Introduction}
All realistic quantum systems interact with an environment and a proper description of their dynamics calls for the toolbox of open quantum systems theory \cite{bp, weiss}. In this framework the simplest type of noise is described by a Markovian master equation in the Lindblad form, corresponding to a completely positive, trace preserving dynamical map satisfying the semi-group property \cite{linblad}. The latter condition means that the map can be divided into infinitely many time-steps, each identical and independent of the past and future steps \cite{wolf}, and therefore a Markovian dynamical map has the intuitive interpretation of memoryless dynamics. Markovian processes successfully describe a plethora of physical processes, particularly in the field on quantum optics, but can fail if applied to more complex system-environment interactions where memory effects become important. In such situations one must resort to non-Markovian dynamical maps.\\
Recently the theory of non-Markovian dynamics has beautifully taken shape as a result of proposals for the definition of non-Markovian dynamics \cite{wolf, BLP, RHP, fisher, cv, luo}. Amazingly, in the past the whole concept has lacked a simple, model-independent definition. The application of non-Markovianity quantifiers, constructed on the basis of these definitions, has led to a deeper understanding of the microscopic mechanisms underlying non-Markovian dynamics \cite{microscopic}, and to the clarification of hazy concepts \cite{hazy}. Moreover, it turns out that the quantifiers can be used to witness initial correlations in the composite system-environment state \cite{initialcorrelations}, and in the environment state, and to probe quantum phase transitions of the environment \cite{probeus}, to name just a few examples. The quantifier put forward by Breuer {\it et al.}, equating non-Markovianity with bidirectional information flow, has also been studied in a linear optics set-up, thus establishing that non-Markovianity quantifiers are not merely a theoretical tool \cite{nmexperiment}.\\
In the spirit of these advances we recently conducted an investigation of the non-Markovian dynamics of a qubit coupled to an ultracold Bose-Einstein condensed (BEC) gas with a two-fold motivation \cite{us}. On one hand, experimentalists have discovered astonishingly accurate means of controlling and manipulating ultracold gases \cite{ultracoldexperiment}. This raises a question whether ultracold gases could provide a tailored environment to a quantum system such that its decohering effect on the system is minimized. Indeed, we discovered that simple and experimentally feasible manipulation of the ultracold reservoir leads to significant changes in the way the qubit dephases and enables a perfect control of the Markovian to non-Markovian transition in the qubit dynamics. On the other hand, the way a qubit decoheres may reveal important information about the environment, leading to the concept of a probe qubit \cite{probequbit, dieter, dieter2}. Here the fundamental question is to what extent one may probe a large, complicated environment by looking at a simple and accessible quantum system that interacts with it. This work aims to dive deeper into these two aspects in the context of qubits embedded in ultracold gases.\\
More specifically, we study the dynamics of two different qubit models, each embedded in an identical environment, namely a BEC gas. The scattering length of the bosons forming the BEC can be controlled using the Feshbach resonances and therefore we have access to an environment that can be chosen to consist of either free or interacting bosonic particles. It is worth stressing that the latter regime is widely unexplored and prototypes of open system models are mainly built on the assumption of an environment of free particles. Intuitively one may expect the interacting environment to have better memory keeping properties than a non-interacting one and this was exactly what we discovered in Ref. \cite{us}: non-Markovian effects take place when the environment is sufficiently strongly interacting. It was left as an open question, however, whether this phenomenon is specific to the model we studied in Ref. \cite{us} or if one can generally associate interacting environments to non-Markovian dynamics. By comparing and contrasting the reduced dynamics of two different qubit models we find the answer to be negative: an interacting environment can induce Markovian dynamics on one qubit architecture and non-Markovian on another.\\
We also address another unresolved question of non-Markovian open quantum systems, namely the connection between the emergence of memory effects and the form of the spectral density function characterizing the dynamical map. Non-Markovianity is often associated to structured spectra. In the case of the Jaynes-Cummings model, for example, a decrease in the width of the Lorentzian spectral density function always leads to a higher degree of non-Markovianity \cite{jaynescummings}. The connection is much more subtle for purely dephasing processes, as shown in Ref. \cite{discord}, where two of us unveil a condition on the form of the spectrum to create non-Markovian dynamics. In this article we study this connection for the rather complex dephasing dynamics induced by the BEC environment and show that, quite unexpectedly, purely Markovian dynamics can arise from spectral density functions with rich structure.

\section{The two models}

\begin{figure}[h]
  \includegraphics[width=0.9\linewidth]{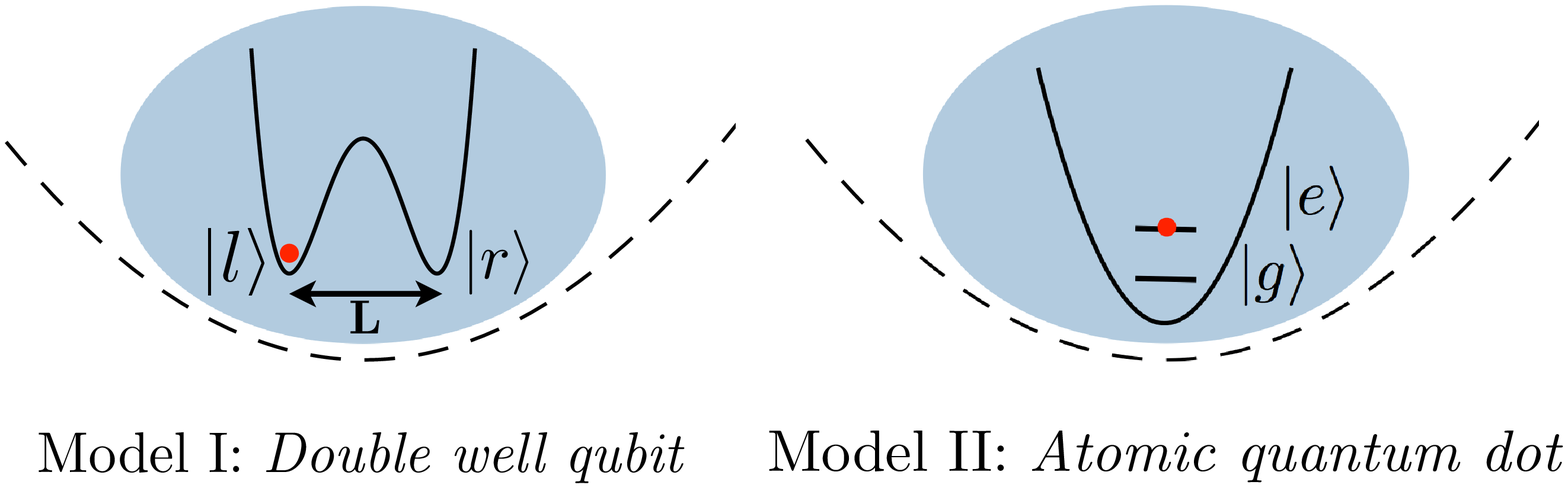}
  \caption{(Color online) The two qubit architectures considered in this article. Model I comprises of an impurity atom trapped in a double well potential and model II assumes a single impurity atom, with an internal level structure, trapped in a deep harmonic trap.  Each qubit interacts with a Bose-Einstein condensed ultracold atomic gas in a shallow harmonic trap.}
  \label{model}
\end{figure}

We compare two different qubit models composed of a trapped impurity atom interacting with an ultracold bosonic gas. In model I, originally introduced in Ref. \cite{massimo} and displayed in Fig. \ref{model}(a), the impurity atom is trapped in a deep double well potential and forms an effective qubit system where the two qubit states are represented by occupation of the impurity in the left $\ket{l}$ or the right well $\ket{r}$. In model II, see Refs. \cite{dieter, dieter2} and Fig. \ref{model}(b), the impurity is trapped in one site of an optical lattice and has two internal states, $\ket{e} \text{and} \ket{g}$, representing the qubit states. The Hamiltonians for both models are composed of three parts: Hamiltonians of the impurity and of the interacting background gas and the interaction Hamiltonian, respectively,
\eqa
&&H_A=\int d\xvec \Psi^\dg(\xvec)\left[\frac{\mathbf{p}_A^2}{2m_A}+V_A(\xvec)\right]\Psi(\xvec),\nn\\
&&H_B=\int d\xvec \Phi^\dg(\xvec)\left[\frac{\mathbf{p}_B^2}{2m_B}+V_B(\xvec)+\frac{g_B}{2}\Phi^\dg(\xvec)\Phi(\xvec)\right]\Phi(\xvec),\nn\\
&&H_{AB}=\frac{g_{AB}}{2}\int d\xvec \Phi^\dg(\xvec)\Psi^\dg(\xvec)\Psi(\xvec)\Phi(\xvec).
\eeqa
Here $m_A$, $\Psi(\xvec)$ and $V_A(\xvec)$ are the mass, field operator and the trapping potential of the impurity atom, $m_B$, $\Phi(\xvec)$, $g_B=4\pi\hbar^2 a_B/m_B$ and $V_B(\xvec)$ are the mass, field operator, coupling constant and the trapping potential of a background gas atom and $a_B$ is the scattering length of the boson-boson collisions. Finally, $g_{AB}=4\pi\hbar^2 a_{AB}/m_{AB}$ is the coupling constant of the impurity-boson interaction where $m_{AB}=m_A m_B/(m_A+m_B)$ is the effective mass.\\
In both models we expand the impurity field operator in terms of Wannier functions $\{\phi_\kvec\}$ localized in the lattice sites/the two wells. Assuming that the lattice sites/the two wells are very deep, hopping and tunneling effects are both suppressed and the Wannier functions take a Gaussian form. We assume that the background gas is weakly interacting and can be treated in the Bogoliubov approximation, neglect all terms that are quadratic in the creation and annihilation operators of the Bogoliubov modes and assume that the background has is homogenous.\\
It turns out that when we focus on a single impurity the Hamiltonians $H_A$ and $H_B$ in moth models are effectively the same. Any differences  in these Hamiltonians will not have an effect on the dynamics of information flow characterising non-Markovian effects (they are all related to the phase of the evolving qubit) and can be safely neglected in this study. The interaction Hamiltonians, instead, have a small but crucial difference, arising from the different trapping potentials of the impurities:
\eqa
H_{AB}^{\text{Model I}}&=&\frac{g_{AB}\sqrt{n_0}}{\Omega}\sum_{\kvec,\;p=L,R}\;\hat{n}_p \hat{c}_k\sqrt{\frac{\epsilon_\kvec}{E_\kvec}}\int d\xvec|\phi(\xvec_p)|^2e^{i \kvec\cdot\xvec}\nn\\
&&\qquad+H.c.,\nn\\
H_{AB}^{\text{Model II}}&=&\frac{g_{AB}\sqrt{n_0}}{\Omega}\sum_\kvec\;\hat{n}\hat{c}_k\sqrt{\frac{\epsilon_\kvec}{E_\kvec}}\int d\xvec|\phi(\xvec)|^2e^{i \kvec\cdot\xvec}\nn\\
&&\qquad+H.c.,
\eeqa
where $n_0$ is the condensate density, $\Omega$ is the quantization volume, $E_\kvec=\sqrt{\epsilon_\kvec(\epsilon_\kvec+2n_0g_B)}$ is the Bogoliubov dispersion relation, $\epsilon_\kvec=\hbar^2k^2/(2 m_B)$ is the dispersion relation of a non-interacting gas with $k=|\kvec|$ and $\hat{c}_k$ is the Bogoliubov excitation operator. Operator $\hat{n}$ is the number operator of the impurities: For model I we assume that there is exactly one impurity atom in the double well system and therefore $\hat{n}_R=\frac{1}{2}(1+\sz)$ and $\hat{n}_L=\frac{1}{2}(1-\sz)$, where $\sz=\ket{l}\bra{l}-\ket{r}\bra{r}$. The two wells are spatially separated by distance $\bf{L}$ so that $\mathbf{x}_R=\mathbf{x}_L-\mathbf{L}$. For model II we also assume one impurity in the lattice site. The atom has one internal state $\ket{g}$ which decouples from the environment and one $\ket{e}$ which does not and therefore $\hat{n}=\ket{e}\bra{e}$.\\
We note that $H_{AB}^{\text{Model I}}$ effectively describes \emph{two} spatially separated qubits of Model II, albeit with a restricted state space $\{\ket{eg}, \ket{ge}\}$. Interestingly these states span the so called subdecoherent state $\ket{eg}+\ket{ge}$, which is very robust against dephasing noise induced by the environment \cite{kalleantti, doll}. Therefore we can expect the qubit architecture of model I to be less affected by noise than model II.\\

\section{Reduced dynamics and information flow} 
The reduced dynamics of both models can be solved analytically \cite{kalleantti, kohler}. Each qubits dephases under the effect of the ultracold gas, i.e., the diagonal elements of the qubit remain constant while the off-diagonals decay as $\rho_{01}(t)=e^{-\Gamma(t)+i\theta(t)}\rho_{01}(0)$. The phase $\theta(t)$ has no effect on the information flow and therefore we do not consider it in this work. Instead we focus on the decoherence function $\Gamma(t)$. When $\Gamma'(t)>0$ information flows from the system to the environment and if there is an interval where $\Gamma'(t)<0$ then the flow of information is temporarily reversed. We associate this reversal with non-Markovian effects, adopting the proposal of Breuer {\it et al.} as our definition for non-Markovianity \cite{BLP}. Indeed, this measure of non-Markovianity,  applied to a dephasing model such as the two models considered here is
\eq \label{measure}
\mathcal{N}=\int_{\Gamma'(t)<0}ds\,\Gamma'(s).
\eeq
Recall that this measure captures the maximal amount of information that can flow back from the environment to the system. The decoherence functions for the two physical systems considered here are
\eqa
\Gamma(t)^{\text{Model I}}&=&\frac{g_{AB}^2n_0}{\Omega}\sum_\kvec e^{-k^2\sigma^2/2}\frac{\epsilon_\kvec}{E_\kvec}\frac{\sin^2(\frac{E_\kvec t}{2\hbar})}{E_\kvec^2}\sin^2\kvec\cdot\mathbf{L},\nn\\
\Gamma(t)^{\text{Model II}}&=&\frac{g_{AB}^2 n_0}{\Omega}\sum_\kvec e^{-k^2\sigma^2/2}\frac{\epsilon_\kvec}{E_\kvec}\frac{\sin^2(\frac{E_\kvec t}{2\hbar})}{E_\kvec^2},
\eeqa
where $\sigma$ is the variance parameter. Interestingly the decoherence factor of Model I has \emph{exactly} the structure of the decoherence factor of two qubits of Model II with spatial separation $\mathbf{L}$ in a subdecoherent state \cite{massimo}, as we anticipated in the previous Section. Therefore we can expect some "coherence trapping" in model I that we would not observe in Model II. 

\begin{figure}[h]
  \includegraphics[width=0.75\linewidth]{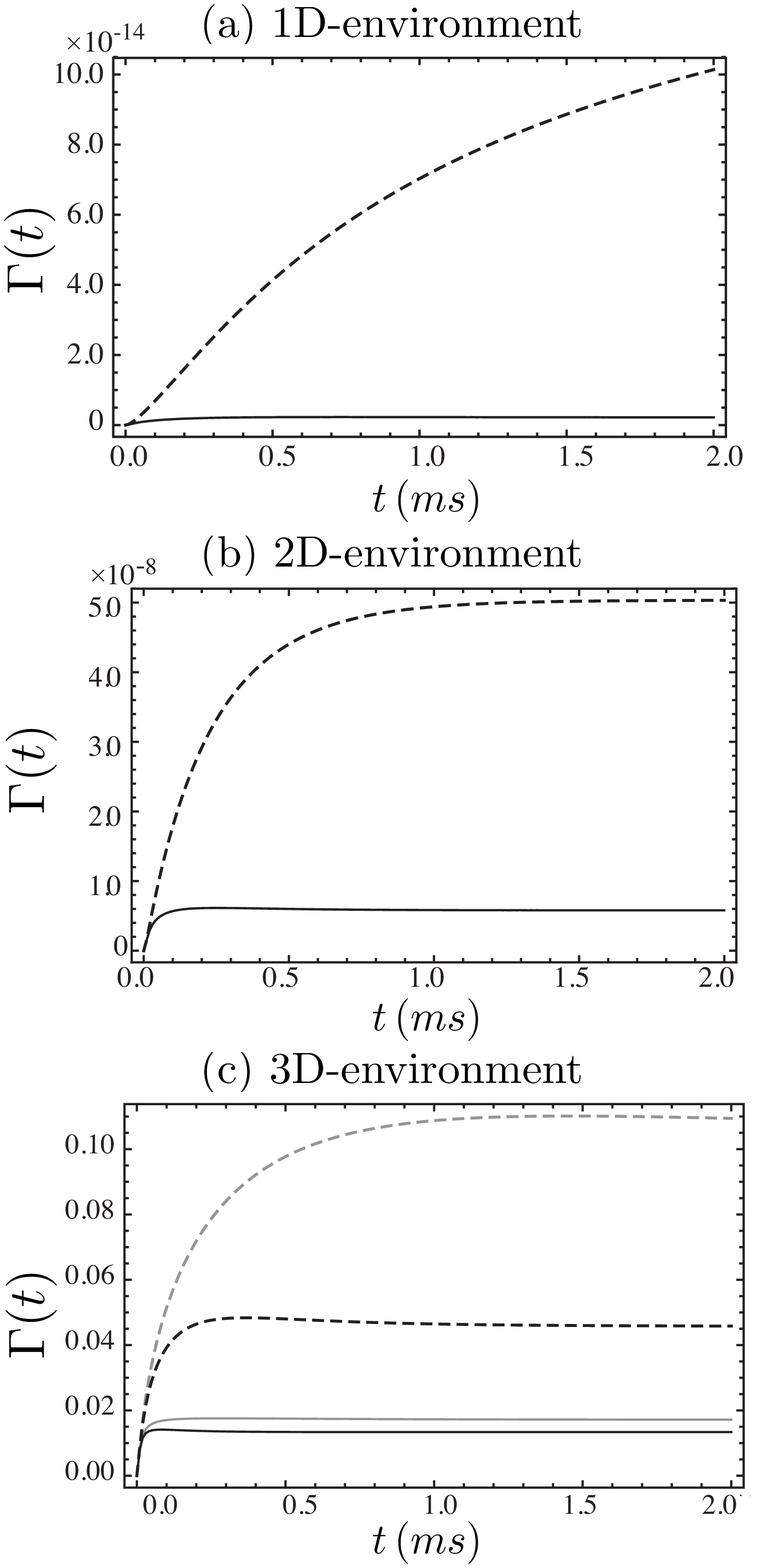}
  \caption{Decoherence functions $\Gamma(t)^{\text{Model I}}$ (solid line) and $\Gamma(t)^{\text{Model II}}$ (dashed line) for  (a) one-dimensional, (b) two-dimensional and (c) three-dimensional environment for $a_B= 0.25 a_{Rb}$ (black lines) and $a_B= a_{Rb}$ (gray lines).}
  \label{decoherence}
\end{figure}

\subsection{Dynamics of the decoherence factor}
We plot the decoherence factors of the two models in Fig. \ref{decoherence}, using the same values of parameters as in Ref. \cite{us} and going to the limit of a continuum of modes, $\Omega^{-1}\sum_\kvec\rightarrow(2\pi)^D\int d\kvec$, where $D$ is the dimension of the BEC. As anticipated, the qubit of Model I is much more robust against decoherence than the qubit of Model II. This difference is most striking in the case of a quasi-1D environment, where $\Gamma(t)^{\text{Model I}}$ saturates quickly to a small value, while $\Gamma(t)^{\text{Model II}}$ increases without bound over all considered time-scales. Hence Model I is almost unaffected by the environmental noise while Model II loses all coherence, and all initial states converge towards the maximally mixed state. The difference in the dynamics of the two models is less drastic when the environment is quasi-2D or 3D, where both decoherence factors saturate to a stationary value, although also in these two cases Model I is more robust against the noise than Model II. \\
Furthermore, the decoherence factor of Model II is monotonic in the case of 1D and 2D background gases, implying that the flow of information is always from to the qubit to the environment and the dynamics is Markovian. This is at variance with the decoherence of Model I where we find both Markovian and non-Markovian dynamics in the case of 1D and 2D background gases, depending on the value of the scattering length of the background gas \cite{us}. In the case of a 3D background gas, shown in Fig. \ref{decoherence}(c), both decoherence factors are non-monotonic for a large enough value of the scattering length. This signals non-Markovian effects. In the next Section we quantify these using the measure of Eq. \eqref{measure}.

\subsection{Non-Markovianity} 
In all cases we have considered we only see one period of information backflow. This permits the use of a slightly modified non-Markovianity measure, which captures the maximal \emph{fraction} of information that can flow back from the environment to the system after an initial period of information flowing from the system to the environment \cite{us}. In Fig. \ref{measure} we show this modified non-Markovianity measure against the scattering length of the background gas in the case when the qubits are immersed in a 3D background gas. We observe a crossover from Markovian to non-Markovian dynamics for both models, although with different values of the crossover point: for a range of scattering lenghts Model I decoheres in a non-Markovian way, while the dynamics of Model II is Markovian. We also consider the effect of temperature on both systems. For a small temperature $T=10$ nK we still observe a Markovian to non-Markovian crossover for Model I, although the crossover point has been shifted slightly to a larger scattering length. Physically this means that in order to induce non-Markovian dynamics, the boson-boson interaction has to be stronger to overcome the detrimental effects of thermal fluctuations.  For Model II the measure is zero (the dynamics is Markovian) for all the considered values of the scattering length, demonstrating the non-Markovianity of Model II is very fragile against thermal effects in the environment. We explain the differences in the dynamics of the two models in the following Section.

\begin{figure}[h]
  \includegraphics[width=0.75\linewidth]{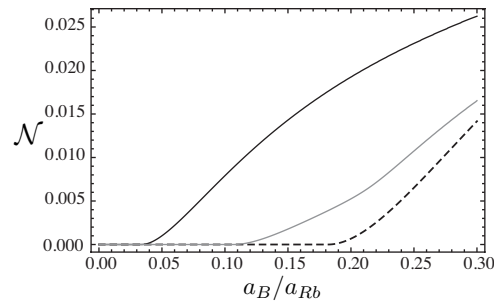}
  \caption{Non-Markovianity measure for model I in a zero-T reservoir (solid black line) and a $T=10$ nK reservoir (solid gray line) and model II in a zero-T reservoir (dashed black line).}
  \label{measure}
\end{figure}

\begin{figure}[h]
  \includegraphics[width=\linewidth]{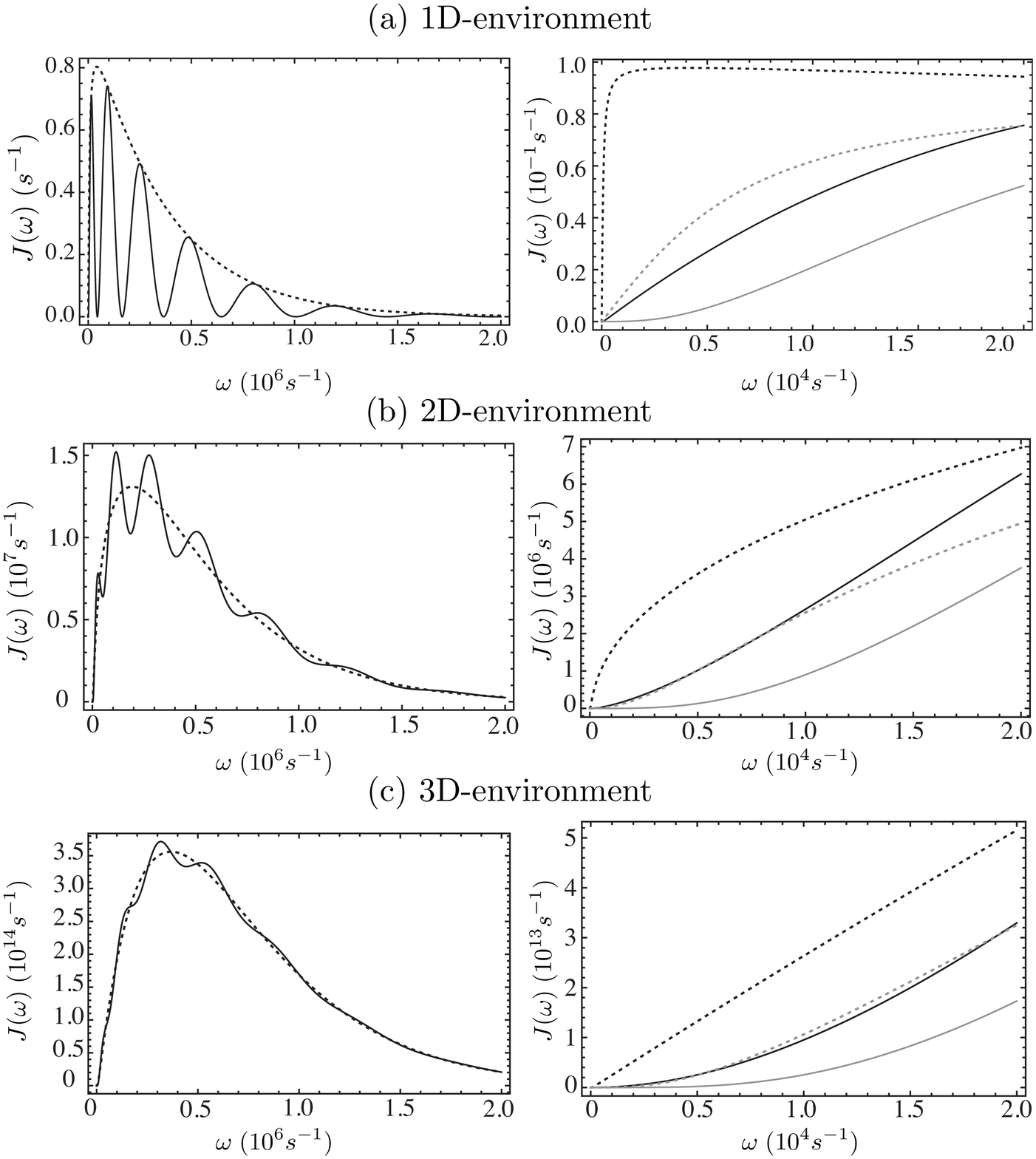}
  \caption{Spectral density functions $J(\omega)$ for Model I (solid lines in all figures) and Model II (dashed lines in all figures) in a (a) one-dimensional, (b) two-dimensional and (c) three-dimensional environment. Left hand side figures show the full spectrum, and the figures on the right show the low-frequency contribution. In the latter we show the spectrum for a weakly interacting background gas with $a_B=10^{-3}a_{Rb}$ (black lines) and for a BEC with $a_B=a_{Rb}$ (gray lines).}
  \label{spectra}
\end{figure}

\section{Spectral density functions}
The spectral density function  $J(\omega)=\sum_\kvec|g_\kvec|^2\delta(\omega-\epsilon_\kvec)$ characterising the dephasing dynamics of an open quantum system is determined by the coupling constants $g_\kvec$ of the effective interaction Hamiltonian $H_{AB}=\sigma_z\sum_\kvec g_\kvec b_\kvec^\dagger+H.c.$ The spectrum of each qubit model considered in this article provides a framework for explaining the notable differences in their dynamics, namely the Markovian to non-Markovian crossover and the increased robustness against environmental noise of Model I compared to Model II. The spectral density, for small frequencies, shows a power-law behaviour $J(\omega)\propto \omega^s$. In the following we call parameter $s>0$ the Ohmicity parameter since it determines whether the spectral density is Ohmic with $s=1$, sub-Ohmic with $s<1$ or super-Ohmic with $s>1$.\\
We conjectured in Ref. \cite{us} that the existence of the crossover from Markovian to non-Markovian dynamics is closely related to which class of the Ohmic spectral densities (sub-Ohmic, Ohmic or super-Ohmic) the spectrum belongs to. In Ref. \cite{discord} we further quantified this claim, presenting a necessary condition for non-Markovian dephasing dynamics: Non-Markovian dynamics can only appear if the spectrum of the environment is super-Ohmic. More specifically, we showed that for a very general dephasing model introduced by Palma, Suominen and Eckert, with a spectrum of the form $J(\omega)=\eta\;\omega_{ph}^{1-s}\omega^s\exp\{-\omega/\omega_c\}$, where $\eta$ is a dimensionless coupling constant that we set to unity and $\omega_{ph}$ is a phononic reference frequency introduced to keep the dimension of the spectrum equal for all values of $s$) \cite{weiss}, non-Markovian effects take place only if $s>s_{crit}=2$, i.e., the spectrum is strongly super-Ohmic. For the physical models considered here the form of the spectral density function is more complicated, but we show that the main results hold also for these two systems.

\subsection{Spectra at low frequencies}
We plot the spectral densities $J(\omega)$ for the two models in Fig. \ref{spectra}, focusing on the low-frequency part of the spectum in the right hand side column. We observed in Ref. \cite{us} that for the ultracold environments the effective Ohmicity parameter depends on the scattering length of the environmental bosons, $s=s(a_B)$ and on the dimensionality of the BEC environment. Changing these two allows transitions from one Ohmic class to another. Here we confirm that for both models and for all three dimensions increasing the scattering length increases the effective value of $s$ for small frequencies $\omega$. This effect is especially pronounced in the case of Model II. When the environment is a quasi-1D free gas ($a_B/a_{Rb}\ll1$) the low-frequency part of the spectrum is sub-Ohmic but as the scattering length is increased to $a_B/a_{Rb}\approx1$ ,the spectrum approaches an Ohmic form. However, even with further increase in the strength of the boson-boson coupling in the environment the spectrum does not become super-Ohmic and indeed we never observe non-Markovian effects in Model II in the 1D regime. In the case of a quasi-2D environment the spectrum of the environment changes from sub-Ohmic to super-Ohmic when the scattering length value is increased, but even in the more strongly interacting case the spectrum is not sufficiently super-Ohmic to trigger non-Markovian effects. Instead in the 3D case the free gas has an Ohmic spectrum which turns super-Ohmic as the interaction between the bosons is turned on and increased. In this case the spectrum can become so super-Ohmic that we observe non-Markovian effects.\\
Model I is already naturally more Ohmic than Model II in the sense that when the ultracold gas is essentially non-interacting both the quasi-2D and the 3D environments have a super-Ohmic spectrum (the quasi 1D environment is roughly Ohmic) and the spectra become super-Ohmic in all dimensions when the scattering length is increased. Critically, in all dimensions the spectrum becomes super-Ohmic enough to create non-Markovian effects in the dynamics of the double well qubit for some critical value of the scattering length, and therefore we observe a crossover from Markovian to non-Markovian dynamics for Model I in all three dimensions.

\subsection{Full spectrum}
Changes in the scattering length have a crucial effect on the spectral density for small frequencies but they are almost negligible when looking at the full spectra. Instead the full spectra, shown in the left-hand-side column of Fig.  \ref{spectra}, exhibits another very interesting difference between the two models, namely that the spectrum of Model I oscillates as a function of $\omega$. Moreover, the spectral density function vanishes for some specific values of $\omega$ in the quasi-1D case, implying that some modes of the environment are completely decoupled from the qubit. We find that the larger is the separation between the two wells the more roots the spectral density has, i.e., more modes decouple from the qubit. We observed numerically that increasing the well separation also increases the "coherence trapping", leading to higher stationary values of the off-diagonal elements of the qubit density matrix. The higher is the dimension of the environment, the smaller are the deviations of the spectrum of Model I compared to that of Model II. This phenomenon is also reflected in the differences in the decoherence factors of the two models; the differences are most pronounced in the case of a quasi-1D environment and in the higher dimensions the decoherence factors are more similar in both value and dynamical behaviour.

\section{Discussion and conclusions}
We have studied the non-Markovian dynamics of two physically different realisations of a qubit interacting with a BEC environment. We discovered that the qubit architecture of Model I is much more sensitive to non-Markovian effects than the one used in Model II. This statement applies in three ways: (i) Model I has non-Markovian dynamics in all three dimensions of the BEC, unlike Model II, which is Markovian in 1D and 2D environemnts: (ii) it has larger values of the non-Markovianity measure in the cases when both qubits have non-Markovian dynamics, i.e., it regains previously lost information more easily: (iii) it is more robust against thermal fluctuations. We also discovered that the two qubit architectures can have extremely different sensitivity to environmental noise, especially in the case when the environment is effectively one-dimensional. This result highlights the importance of choosing suitable qubit systems when designing quantum simulations and quantum probe systems.\\
We also explored the connection between the form of the spectral density function and the ensuing qubit dynamics. While the form of the full spectral density function dictates the general dynamics of the qubit, only a very small part of it controls the Markovian to non-Markovian crossover. We noted the importance of the low-frequency part of the spectrum in the emergence of non-Markovian phenomena in Ref. \cite{discord} in the case of a simple dephasing model, and the study presented in this article confirms that the statement holds also for more complex systems. It is nonetheless striking to notice the overwhelming importance of the low-frequency modes. The spectral density function of model I in a quasi-1D BEC has a very rich structure over the frequency range of the order of the cut-off frequency $\sigma^{-1}$, yet this has no effect on the crossover of the qubit dynamics from Markovian to non-Markovian. The crossover is fully controlled by the behaviour of the spectral density function for low frequencies $\omega\ll\sigma^{-1}$, specifically whether the spectrum is quadratically increasing or not. It is worth stressing that this connection seems to be quite specific to pure dephasing noise. In Ref. \cite{nori} the authors studied a dissipative model and found a direct connection between roots of the spectral density function and non-Markovian dissipationless dynamics; here we discovered that in the pure dephasing qubit model the roots of the spectra do not at all affect the (non-)Markovianity of the dynamics.\\
In conclusion, our results have twofold importance. On one hand they illustrate in a very clear way how the connection between non-Markovianity (or memory effects) and structured environments generally has to be taken with great care. On the other hand, and most importantly, they warn us of the misuse of the term "non-Markovian environment". In our study the environment is exactly the same for the two models, and in both cases it is interacting with a qubit probe. However, under certain conditions, perfectly identical environments induce Markovian dynamics on one qubit and non-Markovian dynamics on another.

\begin{acknowledgments}
We acknowledge financial support from EPSRC (EP/J016349/1), the Finnish Cultural Foundation (Science Workshop on Entanglement) and the Emil Aaltonen foundation (Non-Markovian Quantum Information) and Magnus Ehrnroth foundation. We thank Gabriele De Chiara, Dieter Jaksch, Stephen Clark and Tomi Johnson for helpful discussions.
\end{acknowledgments}

\end{document}